\documentclass[10pt,twocolumn,amsmath,amssymb,aps,prl,showpacs,longbibliography,superscriptaddress]{revtex4-1}
\usepackage[latin9]{inputenc}
\usepackage{amsmath}
\usepackage{amssymb}
\usepackage{graphicx}
\usepackage{esint}

\makeatletter
\usepackage{amsfonts}
\usepackage{graphics}

\makeatother

\begin{document}
\title{Theory of the spectral function of Fermi polarons at finite temperature}
\author{Hui Hu}
\affiliation{Centre for Quantum Technology Theory, Swinburne University of Technology,
Melbourne 3122, Australia}
\author{Jia Wang}
\affiliation{Centre for Quantum Technology Theory, Swinburne University of Technology,
Melbourne 3122, Australia}
\author{Xia-Ji Liu}
\affiliation{Centre for Quantum Technology Theory, Swinburne University of Technology,
Melbourne 3122, Australia}
\date{\today}
\begin{abstract}
We develop a general theory of Fermi polarons at nonzero temperature,
including particle-hole excitations of the Fermi sea shake-up to arbitrarily
high orders. The exact set of equations of the spectral function is
derived by using both Chevy ansatz and diagrammatic approach, and
their equivalence is clarified to hold in free space only, with an
unregularized infinitesimal interaction strength. The correction to
the polaron spectral function arising from two-particle-hole excitations
is explicitly examined, for an exemplary case of Fermi polarons in
one-dimensional optical lattices. We find quantitative improvements
at low temperatures with the inclusion of two-particle-hole excitations,
in both polaron energies and decay rates. Our exact theory of Fermi
polarons with arbitrary orders of particle-hole excitations might
be used to better understand the intriguing polaron dynamical responses
in two or three dimensions, whether in free space or within lattices.
\end{abstract}
\maketitle
Fermi polarons, which are quasiparticles describing the collective
motion of an impurity as it interacts with and shakes up a Fermi sea,
manifest in various realms of condensed matter physics \citep{Alexandrov2010}.
This well-established concept underlies a number of fantastic quantum
many-body phenomena, including Anderson orthogonality catastrophe
\citep{Anderson1967}, the Fermi edge singularity in x-ray spectra
\citep{Mahan1967,Nozieres1969}, and Nagaoka ferromagnetism \citep{Nagaoka1966,Shastry1990,Cui2010}.
The recent realization of atomic Fermi gases with spin-population
imbalance opens a new paradigm to quantitatively explore Fermi polaron
physics in untouched territory \citep{Massignan2014,Schmidt2018},
owing to the unprecedented controllability of ultracold atoms \citep{Bloch2008},
particularly in interatomic interactions \citep{Chin2010}. Thus far,
considerable attention has been given to investigating the ground
state of Fermi polarons \citep{Massignan2014}, known as attractive
polarons, through both experimental and theoretical means. The attractive
polaron energy has been calculated to great accuracy, by using methods
such as variational Chevy ansatz \citep{Chevy2006,Combescot2008,Giraud2010},
diagrammatic $T$-matrix approach \citep{Giraud2010,Combescot2007,Hu2018,Mulkerin2019,Tajima2019,Hu2022},
functional renormalization group \citep{Schmidt2011,vonMilczewski2024},
and quantum Monte Carlo simulations \citep{Prokofev2008}. The outcomes
of these predictions align remarkably well with spectroscopic measurements,
including radio-frequency (rf) spectroscopy \citep{Schirotzek2009,Zhang2012,Zan2019},
Ramsey interferometry \citep{Cetina2016}, Rabi cycle \citep{Scazza2017,Vivanco2024},
and Raman spectroscopy \citep{Ness2020}.

In contrast, describing the excited states of Fermi polarons proves
to be notably challenging \citep{Goulko2016}, especially when departing
from the heavy impurity limit, where exact numerical calculations
might be feasible \citep{Schmidt2018,Knap2012,Wang2022PRL,Wang2022PRA}.
As a result, the finite-temperature dynamical responses of Fermi polarons
related to excited states, as assessed by various spectroscopic studies,
are less well understood. Specifically, in the case of unitary Fermi
polarons with an infinitely large scattering length at degenerate
temperature, state-of-the-art diagrammatic $T$-matrix theory \citep{Tajima2019,Hu2022}
falls short in explaining the spectral features observed in the rf
spectroscopy \citep{Zan2019}. These features unveil the abrupt dissolution
of the attractive polaron, leading to the emergence of excited branches
featuring either repulsive polarons or dressed dimerons. Additionally,
the theory struggles to provide a quantitative explanation for the
observed Raman spectra \citep{Hu2022Raman}, when the interaction
between the impurity and Fermi sea becomes strong. The inadequacy
of the theory at nonzero temperature may stem from its insufficient
description of the Fermi sea shake-up, as it only includes one-particle-hole
excitations of the Fermi sea \citep{Combescot2007,Hu2022}.

In this Letter, we present a formally exact finite-temperature theory
of Fermi polarons, incorporating arbitrary numbers of particle-hole
excitations of the Fermi sea. We use two methods to derive an exact
set of equations for the fundamental quantity of the polaron spectral
function, which determines the rf, Ramsey and Raman spectroscopies.
The first method of Chevy ansatz is generally applicable to any interaction
potential, while the second diagrammatic approach is restricted to
a contact interaction in free space, whose unregularized strength
is infinitesimal. Remarkably, Our diagrammatic theory presents a very
rare case that a quantum many-body system can be exactly solved by
finding out the complete series of Feynman diagrams. We establish
the equivalence of the two approaches when they are both valid and
show that the coefficients in Chevy ansatz can be directly expressed
in terms of the many-particle vertex functions in the diagrammatic
theory. A more comprehensive discussion of the derivations of these
two approaches is presented in a companion paper \citep{LongPRA2024}.

The exact set of equation for the spectral function can be truncated
to enclose, to a particular order (i.e., $n$-th order with $n$ particle-hole
excitations). To illustrate, we focus on Fermi polarons in one-dimensional
lattices and analyze the enhanced predictive capabilities of the spectral
function when two-particle-hole excitations are taken into account.
Future studies with more involved numerical efforts would be beneficial
in providing quantitative predictions for the finite-temperature spectral
function of unitary Fermi polarons in three-dimensional free space,
and would offer insights into elucidating the perplexing spectral
features observed in rf and Raman spectroscopies thus far \citep{Zan2019,Ness2020}.

\textit{Chevy ansatz at finite }\textsl{T}. Following the seminal
works \citep{Chevy2006,Combescot2008}, we take the following Chevy
ansatz for a single spin-down atom (i.e., impurity) immersed in a
Fermi sea of spin-up atoms with a total momentum $\mathbf{p}$,
\begin{equation}
\left|\psi\right\rangle =\sum_{n=0}^{\infty}\left|\psi_{n}\right\rangle =\sum_{n=0}^{\infty}\frac{1}{\left(n!\right)^{2}}\sum_{\{\mathbf{k}\mathbf{q}\}}\alpha_{\mathbf{q}_{1}\mathbf{q}_{2}\cdots\mathbf{q}_{n}}^{\mathbf{k}_{1}\mathbf{k}_{2}\cdots\mathbf{k}_{n}}d_{\mathbf{p}-\mathbf{P}_{\vec{\kappa}_{n}}}^{\dagger}\left|\vec{\kappa}_{n}\right\rangle ,
\end{equation}
where $d_{\mathbf{p}}^{\dagger}$ and $c_{\mathbf{k}}^{\dagger}$
are respectively the creation field operators of the impurity and
spin-up atoms, and $\left|\vec{\kappa}_{n}\right\rangle \equiv c_{\mathbf{k}_{1}}^{\dagger}\cdots c_{\mathbf{k}_{n}}^{\dagger}c_{\mathbf{q}_{n}}\cdots c_{\mathbf{q}_{1}}\left|\textrm{FS}\right\rangle $
describes $n$-particle-hole excitations out of a thermal Fermi sea
$\left|\textrm{FS}\right\rangle $, with a momentum $\mathbf{P}_{\vec{\kappa}_{n}}=(\mathbf{k}_{1}+\cdots+\mathbf{k}_{n})-(\mathbf{q}_{1}+\cdots+\mathbf{q}_{n})$.
The occupation of each state $\mathbf{k}$ in the thermal Fermi sea
is given by the Fermi distribution function $f(\xi_{\mathbf{k}})=1/(e^{\xi_{\mathbf{k}}/k_{B}T}+1)$,
where $\xi_{\mathbf{k}}=\varepsilon_{\mathbf{k}}-\mu$ is the dispersion
of spin-up atoms, measured from the chemical potential $\mu$. Due
to the anti-commutation of fermionic field operators, the coefficients
$\alpha_{\mathbf{q}_{1}\mathbf{q}_{2}\cdots\mathbf{q}_{n}}^{\mathbf{k}_{1}\mathbf{k}_{2}\cdots\mathbf{k}_{n}}$
are antisymmetric upon exchanging $\mathbf{k}_{i}$ and $\mathbf{k}_{j}$
or $\mathbf{q}_{i}$ and $\mathbf{q}_{j}$, where $i,j=1,\cdots,n$.
The resulting redundancy is removed by the factor $1/(n!)^{2}$.

We solve the Schrödinger equation, $\mathcal{H}\left|\psi\right\rangle =(\mathcal{H}_{0}+\mathcal{H}_{\textrm{int}})\left|\psi\right\rangle =E\left|\psi\right\rangle $,
based on a crucial observation that $\mathcal{H}\left|\psi_{n}\right\rangle $
can be expressed by a combination of the terms $d_{\mathbf{p}-\mathbf{P}_{\vec{\kappa}_{m}}}^{\dagger}\left|\vec{\kappa}_{m}\right\rangle $,
where $m=n-1$, $n$, and $n+1$, so we may directly write down a
set of equations for the coefficients. The action of the non-interacting,
kinetic part of the Hamiltonian on the wavefunction is easy to work
out \citep{LongPRA2024}, $\mathcal{H}_{0}\left|\psi_{n}\right\rangle =1/(n!)^{2}\sum_{\{\mathbf{k}\mathbf{q}\}}(E_{\textrm{FS}}+\varepsilon_{\mathbf{p}-\mathbf{P}_{\vec{\kappa}_{n}}}^{I}+E_{\vec{\kappa}_{n}})\alpha_{\mathbf{q}_{1}\mathbf{q}_{2}\cdots\mathbf{q}_{n}}^{\mathbf{k}_{1}\mathbf{k}_{2}\cdots\mathbf{k}_{n}}d_{\mathbf{p}-\mathbf{P}_{\vec{\kappa}_{n}}}^{\dagger}\left|\vec{\kappa}_{n}\right\rangle $,
where $E_{\textrm{FS}}$ is the energy of the thermal Fermi sea, $E_{\vec{\kappa}_{n}}=(\varepsilon_{\mathbf{k}_{1}}+\cdots+\varepsilon_{\mathbf{k}_{n}})-(\varepsilon_{\mathbf{q}_{1}}+\cdots+\varepsilon_{\mathbf{q}_{n}})$
is the excitation energy of $n$ particles and holes, and $\varepsilon_{\mathbf{p}}^{I}$
is the impurity dispersion relation. The action of the interaction
Hamiltonian on $\left|\psi_{n}\right\rangle $ is also straightforward
to obtain, after some tedious algebra \citep{LongPRA2024}. For the
simple case of a contact interaction (in free space) or an on-site
interaction (in lattices) with strength $U$, we find\begin{widetext}
\begin{eqnarray}
-E_{\mathbf{p};\{\mathbf{k}\};\{\mathbf{q}\}}^{(n)}\alpha_{\mathbf{q}_{1}\mathbf{q}_{2}\cdots\mathbf{q}_{n}}^{\mathbf{k}_{1}\mathbf{k}_{2}\cdots\mathbf{k}_{n}} & = & U\sum_{i,j=1,\cdots,n}\left(-1\right)^{i+j}\alpha_{\mathbf{q}_{1}\cdots\mathbf{q}_{n-j}\mathbf{q}_{n-j+2}\cdots\mathbf{q}_{n}}^{\mathbf{k}_{1}\cdots\mathbf{k}_{n-i}\mathbf{k}_{n-i+2}\cdots\mathbf{k}_{n}}+U\left[\sum_{\mathbf{K}}\left(\alpha_{\mathbf{q}_{1}\mathbf{q}_{2}\cdots\mathbf{q}_{n}}^{\mathbf{K}\mathbf{k}_{2}\cdots\mathbf{k}_{n}}+\cdots+\alpha_{\mathbf{q}_{1}\mathbf{q}_{2}\cdots\mathbf{q}_{n}}^{\mathbf{k}_{1}\cdots\mathbf{k}_{n-1}\mathbf{K}}\right)f\left(-\xi_{\mathbf{K}}\right)\right.\nonumber \\
 &  & \left.-\sum_{\mathbf{Q}}\left(\alpha_{\mathbf{Q}\mathbf{q}_{2}\cdots\mathbf{q}_{n}}^{\mathbf{k}_{1}\mathbf{k}_{2}\cdots\mathbf{k}_{n}}+\cdots+\alpha_{\mathbf{q}_{1}\cdots\mathbf{q}_{n-1}\mathbf{Q}}^{\mathbf{k}_{1}\mathbf{k}_{2}\cdots\mathbf{k}_{n}}\right)f\left(\xi_{\mathbf{Q}}\right)\right]+U\sum_{\mathbf{KQ}}\alpha_{\mathbf{q}_{1}\mathbf{q}_{2}\cdots\mathbf{q}_{n}\mathbf{Q}}^{\mathbf{k}_{1}\mathbf{k}_{2}\cdots\mathbf{k}_{n}\mathbf{K}}f\left(-\xi_{\mathbf{K}}\right)f\left(\xi_{\mathbf{Q}}\right),\label{eq:ChevyAnsatzSolution}
\end{eqnarray}
\end{widetext}where $E_{\mathbf{p};\{\mathbf{k}\};\{\mathbf{q}\}}^{(n)}\equiv-(E-E_{\textrm{FS}}-\nu U)+\varepsilon_{\mathbf{p}-\mathbf{P}_{\vec{\kappa}_{n}}}^{I}+E_{\vec{\kappa}_{n}}$
at the density (or filling factor) $\nu$, and the left-hand-side
of the equation shows the coefficient of $(E-\mathcal{H}_{0}-\nu U)\left|\psi_{n}\right\rangle $.
The three terms on the right-hand-side of the equation come from $(\mathcal{H}_{\textrm{int}}-\nu U)\left|\psi\right\rangle $,
involving a summation over the particle momentum $\mathbf{K}$ or
the hole momentum $\mathbf{Q}$, which carries either a distribution
function $f(-\xi_{\mathbf{K}})=1-f(\xi_{\mathbf{K}})$ or $f(\xi_{\mathbf{Q}})$.
It is easy to see, Eq. (\ref{eq:ChevyAnsatzSolution}) has a nice
hierarchy structure. In particular, once we discard the last term
on the right-hand-side at a given order, the set of equations for
the coefficients $\alpha_{\mathbf{q}_{1}\mathbf{q}_{2}\cdots\mathbf{q}_{n}}^{\mathbf{k}_{1}\mathbf{k}_{2}\cdots\mathbf{k}_{n}}$
encloses.

At zero temperature, where the sharp Fermi surface at the Fermi wavevector
$k_{F}$ separates the momenta $\left|\mathbf{k}_{i}\right|>k_{F}$
and $\left|\mathbf{q}_{i}\right|<k_{F}$, Eq. (\ref{eq:ChevyAnsatzSolution})
was already derived, up to the second order $n=2$ \citep{Combescot2008}
and $n=3$ \citep{Liu2022}. At nonzero temperature, the first-order
truncation of Eq. (\ref{eq:ChevyAnsatzSolution}) was also recently
discussed \citep{Liu2019}. All these studies emphasize that Chevy
ansatz is variational, so their focus is more on some individual many-body
eigenstates of the system. Here, we are interested in attractive or
repulsive polarons, which may consist of a bundle of many-body eigenstates.
The polaron energy at nonzero temperature does not necessarily become
smaller as we increase the order of particle-hole excitations. It
is therefore more useful to describe Fermi polarons using the polaron
Green function. For this purpose, we may take a continuous variable
$\omega\equiv E-E_{\textrm{FS}}-\nu U$ and interpret Eq. (\ref{eq:ChevyAnsatzSolution})
at the leading order, i.e., $(\omega-\varepsilon_{\mathbf{p}}^{I})\alpha_{0}=U\sum_{\mathbf{KQ}}\alpha_{\mathbf{Q}}^{\mathbf{K}}f\left(-\xi_{\mathbf{K}}\right)f\left(\xi_{\mathbf{Q}}\right)$,
as the condition for the poles of the polaron Green function. Indeed,
we are free to take an un-normalized ansatz with $\alpha_{0}=1$ and
consequently identify the polaron self-energy,
\begin{equation}
\Sigma\left(\mathbf{p},\omega\right)=U\sum_{\mathbf{KQ}}\alpha_{\mathbf{Q}}^{\mathbf{K}}f\left(-\xi_{\mathbf{K}}\right)f\left(\xi_{\mathbf{Q}}\right).\label{eq:RelationSelfEnergy}
\end{equation}
We will soon rigorously examine this identification by using the diagrammatic
theory. Thus, for a given $\mathbf{p}$ and $\omega$, if we are able
to solve the set of Eq. (\ref{eq:ChevyAnsatzSolution}) truncated
to a particular order $n$, we may directly calculate the polaron
Green function $G_{\downarrow}(\mathbf{p},\omega)$ and hence the
spectral function $A(\mathbf{p},\omega)=-\textrm{Im\ensuremath{G_{\downarrow}}(\ensuremath{\mathbf{p}},\ensuremath{\omega})}/\pi$.

\textit{Chevy ansatz with $U=0^{-}$}. In free space and in two or
three dimensions, the contact interaction should be regularized by
using an $s$-wave scattering length. Formally, the interaction strength
$U$ becomes infinitesimal, in order to remove the ultraviolet divergence
at large momentum. In this situation, in Eq. (\ref{eq:ChevyAnsatzSolution})
the terms involving a summation over $\mathbf{Q}$ vanish and we may
simplify the equation, by defining the variables, $G_{\mathbf{q}_{1}\mathbf{q}_{2}\cdots\mathbf{q}_{n}}^{\mathbf{k}_{1}\mathbf{k}_{2}\cdots\mathbf{k}_{n-1}}\equiv U\sum_{\mathbf{K}}\alpha_{\mathbf{q}_{1}\mathbf{q}_{2}\cdots\mathbf{q}_{n}}^{\mathbf{k}_{1}\cdots\mathbf{k}_{n-1}\mathbf{K}}f\left(-\xi_{\mathbf{K}}\right)$.
It is then straightforward to derive the following set of equations
\citep{LongPRA2024},\begin{widetext} 
\begin{eqnarray}
G_{\mathbf{q}_{1}\mathbf{q}_{2}\cdots\mathbf{q}_{n}}^{\mathbf{k}_{1}\mathbf{k}_{2}\cdots\mathbf{k}_{n-1}} & = & \left[\frac{1}{U}+\sum_{\mathbf{K}}\frac{f\left(-\xi_{\mathbf{K}}\right)}{E_{\mathbf{p};\mathbf{k}_{1}\mathbf{k}_{2}\cdots\mathbf{K};\mathbf{q}_{1}\mathbf{q}_{2}\cdots\mathbf{q}_{n}}^{(n)}}\right]^{-1}\left[\sum_{j=1}^{n}(-1)^{j-1}\alpha_{\mathbf{q}_{1}\mathbf{q}_{2}\cdots\mathbf{q}_{n-j}\mathbf{q}_{n-j+2}\cdots\mathbf{q}_{n}}^{\mathbf{k}_{1}\mathbf{k}_{2}\cdots\mathbf{k}_{n-1}}+\right.\nonumber \\
 &  & \left.\sum_{\mathbf{K}}\frac{\sum_{i=1}^{n-1}G_{\mathbf{q}_{1}\mathbf{q}_{2}\cdots\mathbf{q}_{n}}^{\mathbf{k}_{1}\mathbf{k}_{2}\cdots\mathbf{k}_{n-i-1}\mathbf{K}\mathbf{k}_{n-i+1}\cdots\mathbf{k}_{n-1}}}{E_{\mathbf{p};\mathbf{k}_{1}\mathbf{k}_{2}\cdots\mathbf{K};\mathbf{q}_{1}\mathbf{q}_{2}\cdots\mathbf{q}_{n}}^{(n)}}f\left(-\xi_{\mathbf{K}}\right)-\sum_{\mathbf{K}\mathbf{Q}}\frac{G_{\mathbf{q}_{1}\mathbf{q}_{2}\cdots\mathbf{q}_{n}\mathbf{Q}}^{\mathbf{k}_{1}\mathbf{k}_{2}\cdots\mathbf{k}_{n-1}\mathbf{K}}}{E_{\mathbf{p};\mathbf{k}_{1}\mathbf{k}_{2}\cdots\mathbf{K};\mathbf{q}_{1}\mathbf{q}_{2}\cdots\mathbf{q}_{n}}^{(n)}}f\left(-\xi_{\mathbf{K}}\right)f\left(\xi_{\mathbf{Q}}\right)\right],\label{eq:ChevyAnsatzG_U0}
\end{eqnarray}
\end{widetext}which are manifestly antisymmetric with respect to
the exchange of two momenta in $\mathbf{k}_{i}$ or $\mathbf{q}_{i}$.
As we shall see, these seemingly complicated equations have an elegant
explanation in terms of Feynman diagrams.

\textit{Diagrammatic theory}. To this aim, let us introduce the ($n+1$)-particle
vertex function $\Gamma_{n+1}(\{k_{l}\};p,\{q{}_{l}\})$, which describes
the in-medium scatterings among $n$ spin-up atoms in the Fermi sea
and the impurity. The collective notation $\{k_{l}\}$ stands for
$k_{1}k_{2}\cdots k_{n}$, where the incoming four-momentum $k_{l}\equiv(\mathbf{k}_{l},i\omega_{l})$
and $\omega_{l}\equiv(2m_{l}+1)\pi k_{B}T$ is the fermionic Matsubara
frequency with integer $m_{l}$. The same notation is similarly taken
for the out-going momenta \{$q_{l}$\}. We require the spin-up atom
with the incoming four-momentum $k_{n}$ interacts first with the
impurity. While it is not so obvious at this point, the vertex function
$\Gamma_{n+1}$ does not depend on $k_{n}$ when $n\geq2$ \citep{LongPRA2024}.
As such, $\Gamma_{n+1}$ is antisymmetric when we exchange any two
momenta in $\{k_{l}\}_{l\neq n}$ or $\{q_{l}\}$.

\begin{figure}[b]
\begin{centering}
\includegraphics[width=0.5\textwidth]{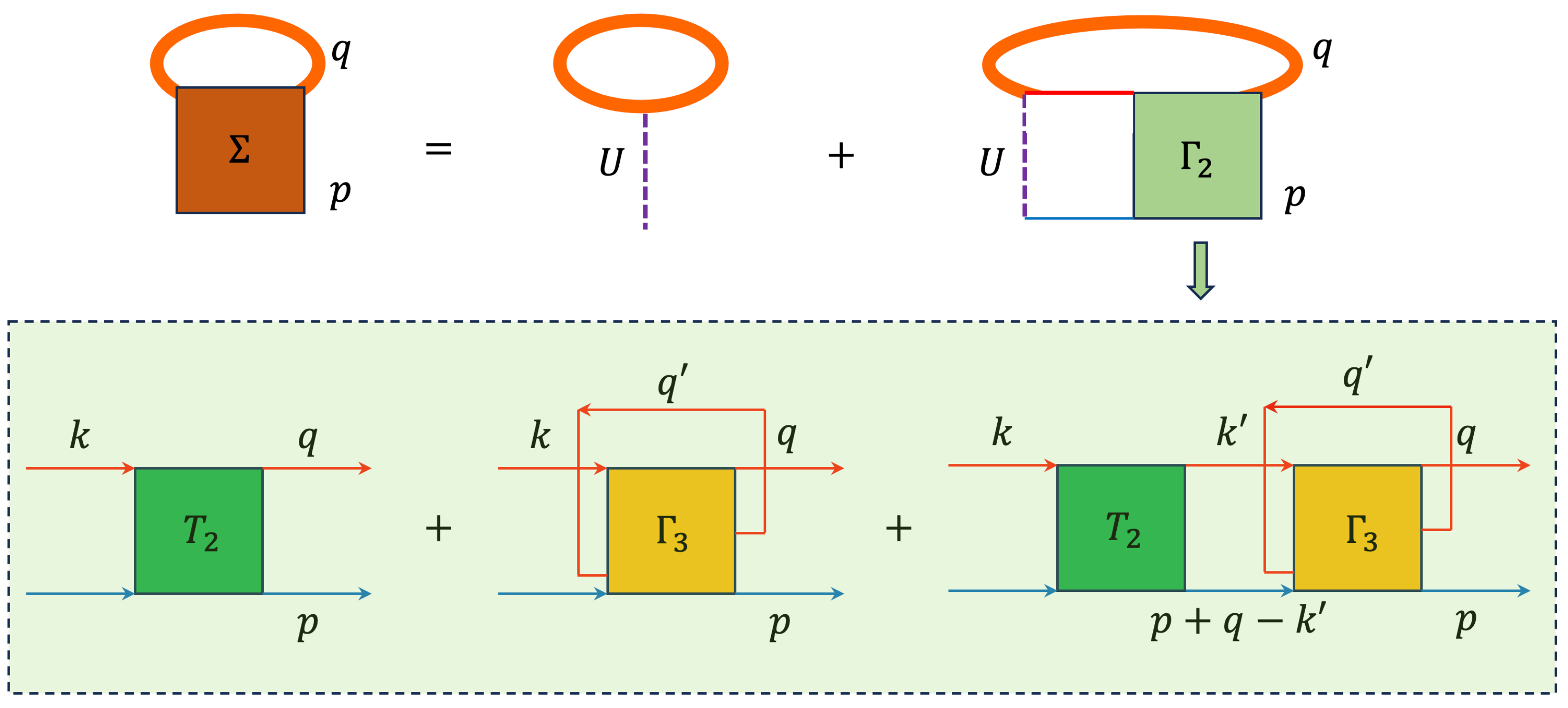}
\par\end{centering}
\caption{\label{fig1: gamma} Polaron self-energy $\Sigma(p)$ expressed in
terms of the vertex function $\Gamma_{2}(k;p,q)$ (see the upper panel),
whose diagrammatic contributions are explicitly listed in the lower
panel, with building blocks of the $T$-matrix $T_{2}(p+q)$ and the
three-body vertex function $\Gamma_{3}(kk';p,qq')$.}
\end{figure}

We find that the coefficients in the Chevy ansatz are related to the
many-particle vertex functions $\Gamma_{n+1}$ \citep{LongPRA2024},
\begin{equation}
\alpha_{\mathbf{q}_{1}\mathbf{q}_{2}\cdots\mathbf{q}_{n}}^{\mathbf{k}_{1}\mathbf{k}_{2}\cdots\mathbf{k}_{n}}=-\frac{\Gamma_{n+1}\left(\{k_{l}\}_{l\neq n};p,\{q_{l}\}\right)}{E_{\mathbf{p};\{\mathbf{k}\};\{\mathbf{q}\}}^{(n)}},\label{eq:RelationAlpha}
\end{equation}
where all the four momenta take the on-shell values, such as $p\equiv(\mathbf{p},\omega)$,
$k_{i}\equiv(\mathbf{k}_{i},\xi_{\mathbf{k}_{i}})$ and $q_{i}=(\mathbf{q}_{i},\xi_{\mathbf{q}_{i}})$.
By integrating over $\mathbf{k}_{n}$ on both sides of the equation
and recalling the fact that $\Gamma_{n+1}$ does not depend on $\mathbf{k}_{n}$,
we obtain $G_{\mathbf{q}_{1}\mathbf{q}_{2}\cdots\mathbf{q}_{n}}^{\mathbf{k}_{1}\mathbf{k}_{2}\cdots\mathbf{k}_{n-1}}=\Gamma_{n+1}\left(\{k_{l}\}_{l\neq n};p,\{q{}_{l}\}\right)$
for $n\geq2$.

\begin{figure}[b]
\begin{centering}
\includegraphics[width=0.5\textwidth]{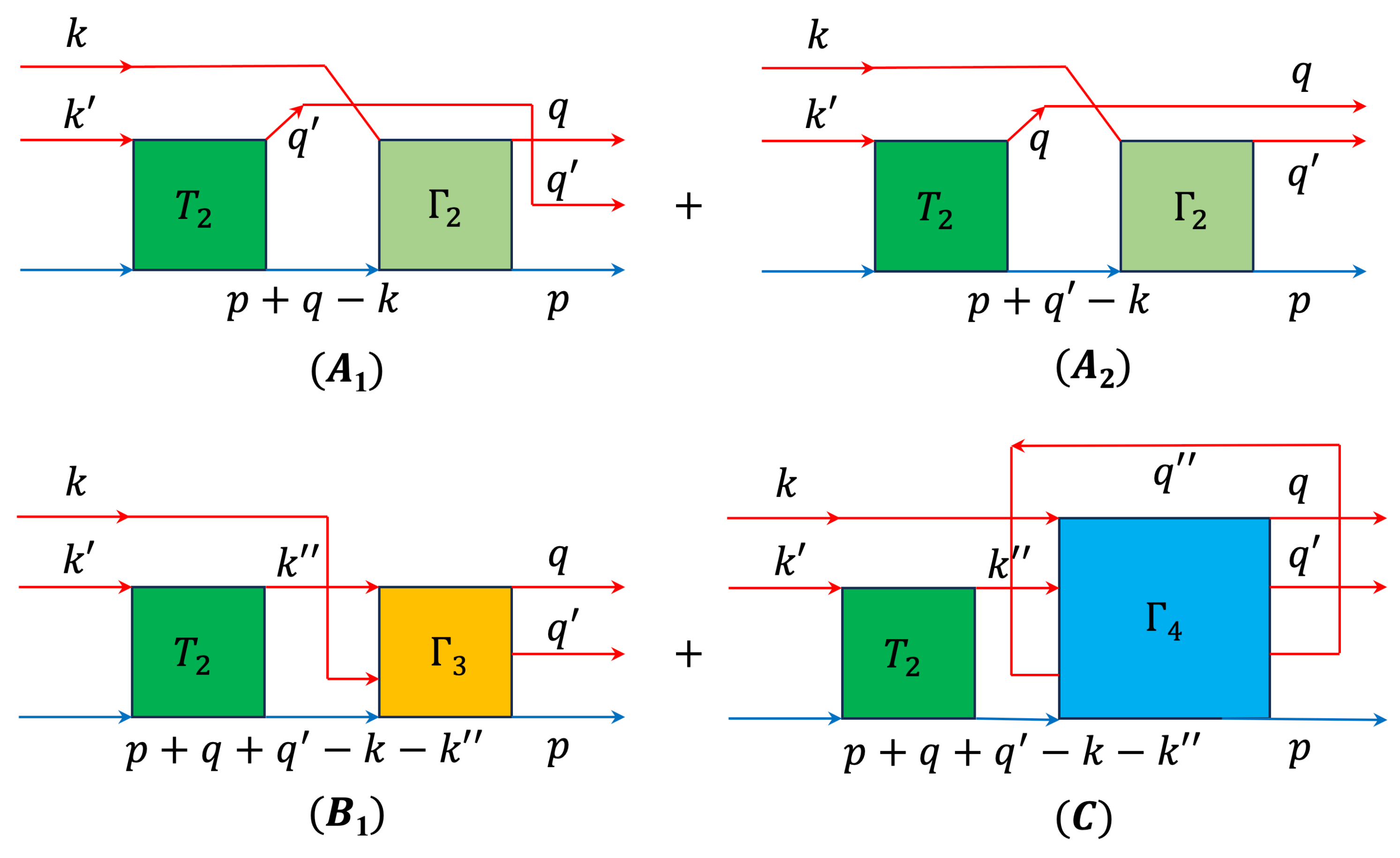}
\par\end{centering}
\caption{\label{fig2: vertex3} Diagrammatic contributions to the three-body
vertex function $\Gamma_{3}(kk';p,qq')$, classified into three different
types of diagrams, $A$, $B$ and $C$, which correspond to the three
terms on the right-hand-side of Eq. (\ref{eq:ChevyAnsatzG_U0}).}
\end{figure}

The relations Eq. (\ref{eq:RelationSelfEnergy}) and Eq. (\ref{eq:RelationAlpha})
are the key results of our Letter, as they clearly demonstrate the
powerfulness of Chevy ansatz in the case of just a few impurities,
which may provide useful insights into further developing accurate
diagrammatic theories for strong correlated systems. To establish
the relations, let us first examine the Dyson equation, which is diagrammatically
shown in Fig. \ref{fig1: gamma}. There, as the impurity line can
only propagate forward \citep{Mahan1967,Nozieres1969}, the vertex
function $\Gamma_{2}$ can be fully represented by three diagrams,
where $T_{2}$ is the standard $T$-matrix that sums up the infinite
ladder diagrams. Similarly, the three-particle vertex function $\Gamma_{3}$
is completely represented by four diagrams as given in Fig. \ref{fig2: vertex3}.
In the companion paper \citep{LongPRA2024}, we also provide the diagrammatic
contributions to the four-particle vertex function $\Gamma_{4}$.

From Fig. \ref{fig2: vertex3}, it is not difficult to write down
the on-shell expression of $\Gamma_{3}$, after we sum over two internal
Matsubara frequencies \citep{LongPRA2024},

\begin{equation}
\frac{\Gamma_{3}\left(k;p,qq'\right)}{T_{2}\left(p+q+q'-k\right)}=A_{1}+A_{2}+B_{1}+C,\label{eq:diagrams2ph}
\end{equation}
where $T_{2}^{-1}(p+q+q'-k)=1/U+\sum_{\mathbf{k}'}f(-\xi_{\mathbf{k}'})/E_{\mathbf{p};\mathbf{kk}';\mathbf{qq}'}^{(2)}$
is the inverse $T$-matrix, and $A_{1}=-\Gamma_{2}(k;p,q)/E_{\mathbf{p};\mathbf{k};\mathbf{q}}^{(1)}=\alpha_{\mathbf{q}}^{\mathbf{k}}$
and $A_{2}=\Gamma_{2}(k;p,q')/E_{\mathbf{p};\mathbf{k};\mathbf{q}'}^{(1)}=-\alpha_{\mathbf{q'}}^{\mathbf{k}}$
are the contributions from the diagrams ($A_{1}$) and ($A_{2}$),
respectively. Finally, the remaining two diagrams give rise to $B_{1}=\sum_{\mathbf{k}'}f(-\xi_{\mathbf{k}'})\Gamma_{3}(k';p,qq')/E_{\mathbf{p};\mathbf{kk}';\mathbf{qq}'}^{(2)}$
and $C=-\sum_{\mathbf{k}'\mathbf{q''}}f(-\xi_{\mathbf{k}'})f(\xi_{\mathbf{q''}})\Gamma_{4}(kk';p,qq'q'')/E_{\mathbf{p};\mathbf{kk}';\mathbf{qq}'}^{(2)}$.
It is easy to check that, in Eq. (\ref{eq:diagrams2ph}) by further
replacing $\Gamma_{3}(k';p,qq')$ by $G_{\mathbf{qq}'}^{\mathbf{k}}$
and $\Gamma_{4}(kk';p,qq'q'')$ by $G_{\mathbf{qq}'\mathbf{q}''}^{\mathbf{kk}'}$,
we indeed recover Eq. (\ref{eq:ChevyAnsatzG_U0}) at the second order
$n=2$. Quite generally, the diagrams of the many-particle vertex
function $\Gamma_{n+1}$ can be categorized into types $A$, $B$
and $C$, which exactly correspond to the three terms on the right-hand-side
of Eq. (\ref{eq:ChevyAnsatzG_U0}), respectively \citep{LongPRA2024}.
The on-shell expression of $\Gamma_{2}$ can be similarly determined
using Fig. \ref{fig1: gamma}. In particular, the Dyson equation reads
\citep{LongPRA2024}, $\Sigma(\mathbf{p},\omega)=-U\sum_{\mathbf{kq}}f(-\xi_{\mathbf{k}})f(\xi_{\mathbf{q}})\Gamma_{2}(k;p,q)/E_{\mathbf{p};\mathbf{k};\mathbf{q}}^{(1)}$,
which is precisely Eq. (\ref{eq:RelationSelfEnergy}), once we use
the relation Eq. (\ref{eq:RelationAlpha}) to replace $-\Gamma_{2}/E_{\mathbf{p};\mathbf{k};\mathbf{q}}^{(1)}$
with $\alpha_{\mathbf{q}}^{\mathbf{k}}$.

\begin{figure}
\begin{centering}
\includegraphics[width=0.5\textwidth]{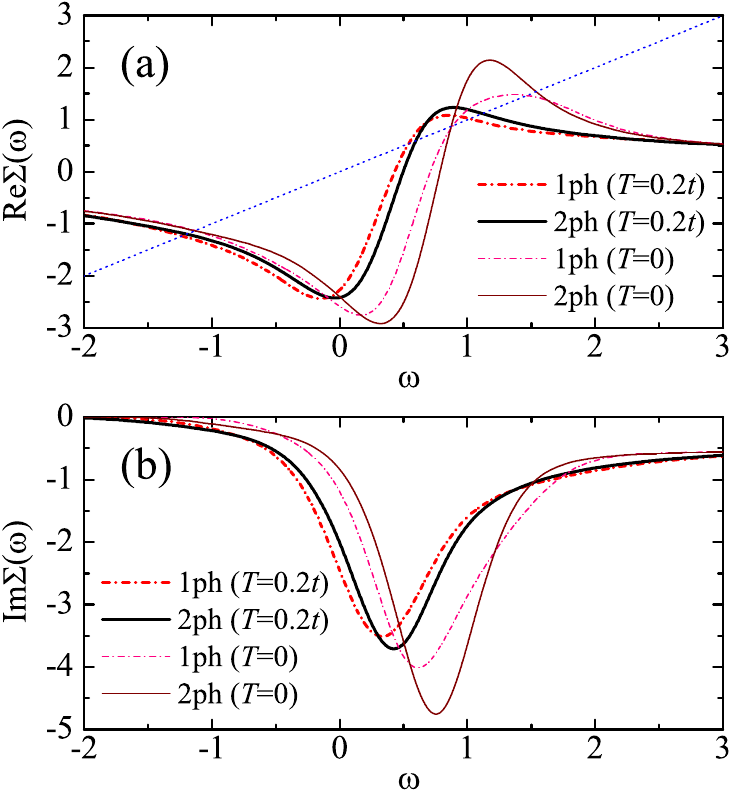}
\par\end{centering}
\caption{\label{fig3: selfenergy} The real part (a) and imaginary part (b)
of the polaron self-energy $\Sigma(\mathbf{p}=0,\omega)$ at zero
momentum. The solid lines and dash-dotted lines correspond to the
results with and without two-particle-hole (2ph) excitations, respectively.
The blue dotted line in (a) shows $\omega-E_{\mathbf{p}=0}$ and its
crossing points with $\textrm{Re}\Sigma$ give rise to the polaron
energies. Here, we take the filling factor $\nu=0.2$, the on-site
interaction strength $U=-4t$, and an impurity hopping amplitude $t_{d}=t$.
Both the self-energy $\Sigma$ and the frequency $\omega$ are measured
in units of the hopping amplitude of atoms $t$.}
\end{figure}

\begin{figure}
\begin{centering}
\includegraphics[width=0.5\textwidth]{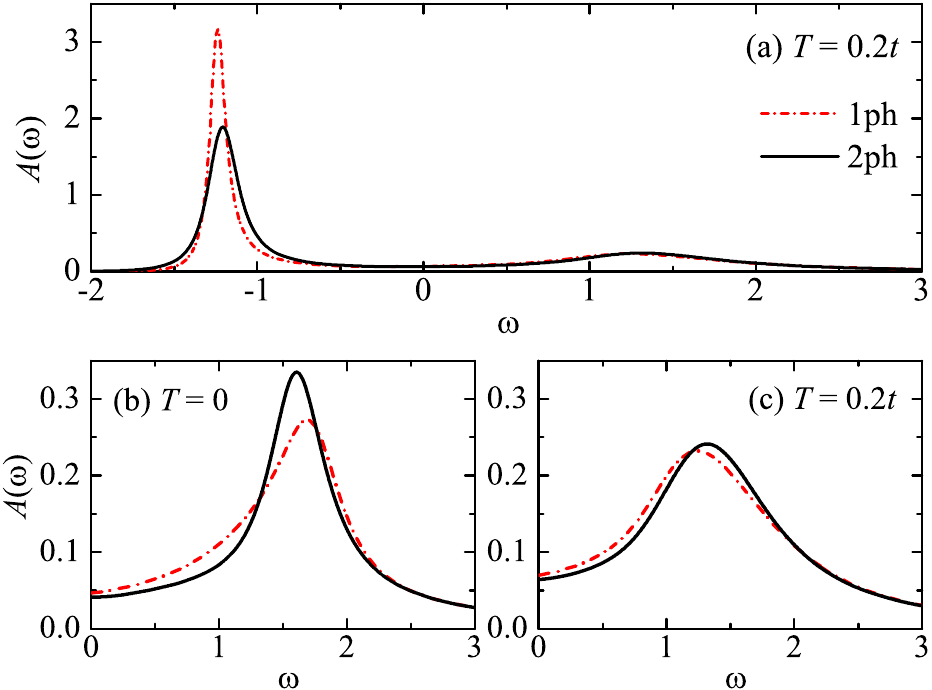}
\par\end{centering}
\caption{\label{fig4: akw} (a) The zero-momentum polaron spectral function
$A(\mathbf{p}=0,\omega)$ at the temperature $T=0.2t$. (b) and (c)
highlight the repulsive polaron responses at $\omega\sim1.5t$, at
zero temperature $T=0$ and at $T=0.2t$, respectively. The black
solid lines or red dash-dotted lines show the predictions with or
without two-particle-hole excitations. The other parameters are the
same as in Fig. \ref{fig3: selfenergy}.}
\end{figure}

\textit{Fermi polarons in lattices}. The exact sets of Eq. (\ref{eq:ChevyAnsatzSolution})
and Eq. (\ref{eq:ChevyAnsatzG_U0}) could be implemented to calculate
the polaron self-energy in Eq. (\ref{eq:RelationSelfEnergy}) and
hence the polaron spectral function. However, numerical calculations
at finite temperature are challenging, due to the zeros of $E_{\mathbf{p};\{\mathbf{k}\};\{\mathbf{q}\}}^{(n)}$
that make the coefficients $\alpha_{\mathbf{q}_{1}\mathbf{q}_{2}\cdots\mathbf{q}_{n}}^{\mathbf{k}_{1}\mathbf{k}_{2}\cdots\mathbf{k}_{n}}$
and $G_{\mathbf{q}_{1}\mathbf{q}_{2}\cdots\mathbf{q}_{n}}^{\mathbf{k}_{1}\mathbf{k}_{2}\cdots\mathbf{k}_{n-1}}$
highly singular. As a result, the truncation to one-particle-hole
excitations was only recently considered \citep{Mulkerin2019,Tajima2019,Hu2022,Liu2019}.
Further improvements to the level of two-particle-hole excitations
have never been attempted.

Here, we focus on Fermi polarons in one-dimensional lattices with
an on-site attraction $U<0$, a situation that can be readily realized
in cold-atom experiments. We solve Eq. (\ref{eq:ChevyAnsatzSolution})
with the inclusion of two-particle-hole excitations \citep{LongPRA2024}.
The singularities in the coefficients are removed by introducing a
finite broadening factor $\eta$ to the frequency, i.e., $\omega\rightarrow\omega_{\eta}\equiv\omega+i\eta$.
We take several broadening factors at the order of the hopping amplitude
of atoms $t$, and eventually extrapolate to $\eta=0^{+}$.

In Fig. \ref{fig3: selfenergy}, we report the polaron self-energy
at zero temperature and at $T=0.2t$, calculated with one-particle-hole
excitations only (red dot-dashed lines) and with two-particle-hole
excitations (black solid lines). We find quantitative improvements
when we incorporate two-particle-hole excitations at both temperatures.
However, the improvement becomes less significant with increasing
temperature. In Fig. \ref{fig4: akw}(a), we present the polaron spectral
function at $T=0.2t$, which clearly shows the attractive branch (at
$\omega\sim-1.3t$) and repulsive branch (at $\omega\sim1.5t$). The
inclusion of two-particle-hole excitations leads to a larger decay
rate for the attractive polaron and thereby a reduced attractive polaron
peak. It also slightly increases attractive polaron energy. In contrast,
for the repulsive polaron, two-particle-hole excitations enhance the
peak height, as revealed by Fig. \ref{fig4: akw}(c). This enhancement
is particularly evident at zero temperature, as shown in Fig. \ref{fig4: akw}(b).

\textit{Conclusions}. In summary, we have derived an exact set of
equations, to determine the spectral function of Fermi polarons, by
using both Chevy ansatz and the diagrammatic approach. Our exact theory
incorporates arbitrary numbers of particle-hole excitations, allowing
a systematic check of the importance of particle-hole excitations
at different level. We have calculated the spectral function of Fermi
polarons in one-dimensional lattices and have examined the improvement
due to the inclusion of two-particle-hole excitations. The extension
of our calculations to a unitary Fermi polaron, with more elaborate
numerical efforts, might be used to quantitatively understand the
puzzling spectral feature observed in recent measurements \citep{Zan2019,Ness2020}.
Furthermore, our exact formalism is also directly applicable to investigate
the few-body (i.e., $n+1$) bound states, which emerge as the poles
of the many-particle vertex functions $\Gamma_{n+1}$ , both in vacuum
or in the presence of the Fermi sea.
\begin{acknowledgments}
This research was supported by the Australian Research Council's (ARC)
Discovery Program, Grants Nos. DP240101590 (H.H.), FT230100229 (J.W.),
and DP240100248 (X.-J.L.).
\end{acknowledgments}


\begin{thebibliography}{99}
\bibitem{Alexandrov2010}A. S. Alexandrov and J. T. Devreese, Advances
in Polaron Physics (Springer, New York, 2010), Vol. 159.

\bibitem{Anderson1967}P. W. Anderson, Infrared Catastrophe in Fermi
Gases with Local Scattering Potentials, Phys. Rev. Lett. \textbf{18},
1049 (1967).

\bibitem{Mahan1967}G. D. Mahan, Excitons in Metals: Infinite Hole
Mass, Phys. Rev. \textbf{163}, 612 (1967).

\bibitem{Nozieres1969}P. Nozières and C. T. De Dominicis, Singularities
in the X-Ray Absorption and Emission of Metals. III. One- Body Theory
Exact Solution, Phys. Rev. \textbf{178}, 1097 (1969).

\bibitem{Nagaoka1966}Y. Nagaoka, Ferromagnetism in a Narrow, Almost
Half-Filled $s$ Band, Phys. Rev. \textbf{147}, 392 (1966).

\bibitem{Shastry1990}B. S. Shastry, H. R. Krishnamurthy, and P. W.
Anderson, Instability of the Nagaoka ferromagnetic state of the $U=\infty$
Hubbard model, Phys. Rev. B \textbf{41}, 2375 (1990).

\bibitem{Cui2010}X. Cui and H. Zhai, Stability of a fully magnetized
ferromagnetic state in repulsively interacting ultracold Fermi gases,
Phys. Rev. A \textbf{81}, 041602(R) (2010).

\bibitem{Massignan2014}P. Massignan, M. Zaccanti, and G. M. Bruun,
Polarons, dressed molecules and itinerant ferromagnetism in ultracold
Fermi gases, Rep. Prog. Phys. \textbf{77}, 034401 (2014).

\bibitem{Schmidt2018}R. Schmidt, M. Knap, D. A. Ivanov, J.-S. You,
M. Cetina, and E. Demler, Universal many-body response of heavy impurities
coupled to a Fermi sea: a review of recent progress, Rep. Prog. Phys.
\textbf{81}, 024401 (2018).

\bibitem{Bloch2008}I. Bloch, J. Dalibard, and W. Zwerger, Many-body
physics with ultracold gases, Rev. Mod. Phys. \textbf{80}, 885 (2008).

\bibitem{Chin2010}C. Chin, R. Grimm, P. Julienne, and E. Tiesinga,
Feshbach resonances in ultracold gases, Rev. Mod. Phys. \textbf{82},
1225 (2010).

\bibitem{Chevy2006}F. Chevy, Universal phase diagram of a strongly
interacting Fermi gas with unbalanced spin populations, Phys. Rev.
A \textbf{74}, 063628 (2006).

\bibitem{Combescot2008}R. Combescot and S. Giraud, Normal State of
Highly Polarized Fermi Gases: Full Many-Body Treatment, Phys. Rev.
Lett. \textbf{101}, 050404 (2008).

\bibitem{Giraud2010}S. Giraud, Contribution à la théorie des gaz
de fermions ultrafroids fortement polarisés (PhD Thesis 2010).

\bibitem{Combescot2007}R. Combescot, A. Recati, C. Lobo, and F. Chevy,
Normal State of Highly Polarized Fermi Gases: Simple Many-Body Approaches,
Phys. Rev. Lett. \textbf{98}, 180402 (2007).

\bibitem{Hu2018}H. Hu, B. C. Mulkerin, J. Wang, and X.-J. Liu, Attractive
Fermi polarons at nonzero temperatures with a finite impurity concentration,
Phys. Rev. A \textbf{98}, 013626 (2018).

\bibitem{Mulkerin2019}B. C. Mulkerin, X.-J. Liu, and Hui Hu, Breakdown
of the Fermi polaron description near Fermi degeneracy at unitarity,
Ann. Phys. (N. Y.) \textbf{407}, 29 (2019).

\bibitem{Tajima2019}H. Tajima and S. Uchino, Thermal crossover, transition,
and coexistence in Fermi polaronic spectroscopies, Phys. Rev. A \textbf{99},
063606 (2019).

\bibitem{Hu2022}H. Hu and X.-J. Liu, Fermi polarons at finite temperature:
Spectral function and rf spectroscopy, Phys. Rev. A \textbf{105},
043303 (2022).

\bibitem{Schmidt2011}R. Schmidt and T. Enss, Excitation spectra and
rf response near the polaron-to-molecule transition from the functional
renormalization group, Phys. Rev. A \textbf{83}, 063620 (2011).

\bibitem{vonMilczewski2024}J. von Milczewski and R. Schmidt, Momentum-dependent
quasiparticle properties of the Fermi polaron from the functional
renormalization group, arXiv:2312.05318.

\bibitem{Prokofev2008}N. Prokof\textquoteright ev and B. Svistunov,
Fermi-polaron problem: Diagrammatic Monte Carlo method for divergent
sign-alternating series, Phys. Rev. B \textbf{77}, 020408(R) (2008).

\bibitem{Schirotzek2009}A. Schirotzek, C.-H. Wu, A. Sommer, and M.W.
Zwierlein, Observation of Fermi Polarons in a Tunable Fermi Liquid
of Ultracold Atoms, Phys. Rev. Lett. \textbf{102}, 230402 (2009).

\bibitem{Zhang2012}Y. Zhang, W. Ong, I. Arakelyan, and J. E. Thomas,
Polaron-to-Polaron Transitions in the Radio-Frequency Spectrum of
a Quasi-Two-Dimensional Fermi Gas, Phys. Rev. Lett. \textbf{108},
235302 (2012).

\bibitem{Zan2019}Z. Yan, P. B. Patel, B. Mukherjee, R. J. Fletcher,
J. Struck, and M.W. Zwierlein, Boiling a Unitary Fermi Liquid, Phys.
Rev. Lett. \textbf{122}, 093401 (2019).

\bibitem{Cetina2016}M. Cetina, M. Jag, R. S. Lous, I. Fritsche, J.
T. M.Walraven, R. Grimm, J. Levinsen, M. M. Parish, R. Schmidt, M.
Knap, and E. Demler, Ultrafast many-body interferometry of impurities
coupled to a Fermi sea, Science \textbf{354}, 96 (2016).

\bibitem{Scazza2017}F. Scazza, G. Valtolina, P. Massignan, A. Recati,
A. Amico, A. Burchianti, C. Fort, M. Inguscio, M. Zaccanti, and G.
Roati, Repulsive Fermi Polarons in a Resonant Mixture of Ultracold
$^{6}$Li Atoms, Phys. Rev. Lett. \textbf{118}, 083602 (2017).

\bibitem{Vivanco2024}F. J. Vivanco, A. Schuckert, S. Huang, G. L.
Schumacher, G. G. T. Assumpção, Y. Ji, J. Chen, M. Knap, and Nir Navon,
The strongly driven Fermi polaron, arXiv:2308.05746.

\bibitem{Ness2020}G. Ness, C. Shkedrov, Y. Florshaim, O. K. Diessel,
J. von Milczewski, R. Schmidt, and Y. Sagi, Observation of a Smooth
Polaron-Molecule Transition in a Degenerate Fermi Gas, Phys. Rev.
X \textbf{10}, 041019 (2020).

\bibitem{Goulko2016}O. Goulko, A. S. Mishchenko, N. Prokof'ev, and
B. Svistunov, Dark continuum in the spectral function of the resonant
Fermi polaron, Phys. Rev. A \textbf{94}, 051605(R) (2016).

\bibitem{Knap2012}M. Knap, A. Shashi, Y. Nishida, A. Imambekov, D.
A. Abanin, and E. Demler, Time-Dependent Impurity in Ultracold Fermions:
Orthogonality Catastrophe and Beyond, Phys. Rev. X \textbf{2}, 041020
(2012).

\bibitem{Wang2022PRL}J. Wang, X.-J. Liu, and H. Hu, Exact Quasiparticle
Properties of a Heavy Polaron in BCS Fermi Superfluids, Phys. Rev.
Lett. \textbf{128}, 175301 (2022).

\bibitem{Wang2022PRA}J. Wang, X.-J. Liu, and H. Hu, Heavy polarons
in ultracold atomic Fermi superfluids at the BEC-BCS crossover: Formalism
and applications, Phys. Rev. A \textbf{105}, 043320 (2022).

\bibitem{Hu2022Raman}H. Hu and X.-J. Liu, Raman spectroscopy of Fermi
polarons, Phys. Rev. A \textbf{106}, 063306 (2022).

\bibitem{LongPRA2024}H. Hu, J. Wang, and X.-J. Liu, Exact theory
of the finite-temperature spectral function of Fermi polarons with
multiple particle-hole excitations: Diagrammatic theory versus Chevy
ansatz, arXiv:2403.09064.

\bibitem{Liu2022}R. Liu, C. Peng, and X. Cui, Emergence of crystalline
few-body correlations in mass-imbalanced Fermi polarons, Cell Reports
Physical Science \textbf{3}, 100993 (2022).

\bibitem{Liu2019}W. E. Liu, J. Levinsen, and M. M. Parish, Variational
Approach for Impurity Dynamics at Finite Temperature, Phys. Rev. Lett.
\textbf{122}, 205301 (2019).
\end{thebibliography}
\end{document}